\RequirePackage{fix-cm}
\documentclass[twocolumn,epjc3]{svjour3}
\smartqed  % flush right qed marks, e.g. at end of proof
\RequirePackage{graphicx}
\RequirePackage{mathptmx}      % use Times fonts if available on your TeX system

\usepackage{cite}
\usepackage{appendix}
\usepackage{epsfig}
\usepackage{amsmath}
\usepackage{amssymb}
\usepackage{bm}
\usepackage{hyperref}
\usepackage{slashed}
\newcommand{\be} {\begin{equation}}
\newcommand{\ee} {\end{equation}}
\newcommand{\ba} {\begin{eqnarray}}
\newcommand{\ea} {\end{eqnarray}}

\newcommand{\dd}{\textrm{d}}

\usepackage{color}
\definecolor{darkblue}{cmyk}{1,0.3,0,0.2}
\definecolor{violet}{cmyk}{0,1,0,0.2}
\hypersetup{colorlinks, bookmarksnumbered, citecolor=darkblue, linkcolor=darkblue, pdfstartview=FitH, urlcolor=darkblue, linktocpage}

\newcommand{\arXhref}[1]{\href{http://arxiv.org/abs/#1}{#1}}
\journalname{Eur. Phys. J. C}

\begin{document}

\title{Higgs Pseudo Observables and Radiative Corrections}

\author{ Marzia Bordone\thanksref{addr2}, Admir Greljo\thanksref{addr2,addr4}, Gino Isidori\thanksref{addr2,addr3}, David Marzocca\thanksref{addr2}, Andrea Pattori\thanksref{addr2}
}

\institute{Physik-Institut, Universit\"at Z\"urich, CH-8057 Z\"urich, Switzerland 
\label{addr2}
\and
INFN, Laboratori Nazionali di Frascati, I-00044 Frascati, Italy \label{addr3}
\and
Faculty of Science, University of Sarajevo, Zmaja od Bosne 33-35, \\ 71000 Sarajevo, Bosnia and Herzegovina \label{addr4}}

\maketitle

\begin{abstract}
We show how  leading radiative corrections can be implemented in the general 
description of $h\to 4\ell$ decays by means of  Pseudo Observables (PO).
With the inclusion of such corrections, the PO  description of  $h\to 4\ell$  decays 
can be matched to next-to-leading-order electroweak 
calculations both within and beyond the Standard Model  (SM). 
In particular, we demonstrate that with the inclusion of such corrections 
the complete next-to-leading-order Standard Model prediction 
for the $h\to 2e2\mu$  dilepton mass spectrum  is recovered within $1\%$ accuracy. The impact of 
radiative corrections for non-standard PO is also briefly discussed. 
\end{abstract}
%%%%%%%%%%%%%%%%%%%%%%%%%%%%%%%%%%%%%%%%
\label{sec:intro}

%%%%%%%%%%%%%%%%%%%%%%%%%%%%%%%%%%%%%%%%
\section{Introduction}
%%%%%%%%%%%%%%%%%%%%%%%%%%%%%%%%%%%%%%%%
\label{intro}

The decays of the Higgs particle, $h(125)$, can be characterised by a set of Pseudo Observables (PO)
that describes, in great generality, possible deviations  from the Standard Model (SM) in the limit of heavy 
New Physics (NP)~\cite{Gonzalez-Alonso:2014eva}.

The Higgs PO are defined from a momentum expansion of the on-shell electroweak 
Higgs decay amplitudes. More precisely, the PO relevant to $h\to 4\ell$ are defined 
by the momentum expansion around the physical poles 
(due to the exchange of SM electroweak gauge bosons) of the following three-point 
correlation function 
\be
	\langle 0 | {\cal T} \left\{ J_{\ell}^\mu(x),  J_{\ell^\prime}^\nu(y), h(0) \right\} | 0 \rangle~,
	\label{eq:ThreePointFunction}
\ee
where $J_{\ell}^\mu(x)$ are generic leptonic currents.
This expansion encodes in full generality 
the {\em short-distance} contributions to the decay amplitudes 
in extensions of the SM with no new light states~\cite{Gonzalez-Alonso:2014eva}.
However, in order to compare this amplitude decomposition with data, 
also the {\em long-distance} contributions  due to soft and collinear photon emission 
(i.e.~the leading QED radiative corrections) 
must be taken into account.  

Soft and collinear photon emission represents a universal correction factor~\cite{Yennie:1961ad,Weinberg:1965nx}
that can be implemented,  by means of appropriate convolution functions 
(or, equivalently, showering algorithms such as those adopted in PHOTOS~\cite{Davidson:2010ew},  PYTH\-IA~\cite{Sjostrand:2007gs}, or 
SHERPA~\cite{Gleisberg:2008ta}) irrespective of the specific short-distance structure of the amplitude.\footnote{For a discussion about the 
implementation of universal QED  corrections in a general EFT context see also Ref.~\cite{Isidori:2007zt}.}
In this paper we illustrate how this works, in practice, in the $h\to 4\ell$ case.

We focus our analysis to the $h\to 2e 2\mu$ case, that is particularly interesting for illustrative purposes:
 the effect of radiative corrections can be implemented by simple analytic 
formulae, allowing a transparent comparison with numerical methods.
As we will show, the inclusion of the universal QED corrections is necessary 
and sufficient to reach an accurate 
theoretical description of the Higgs decay spectrum, that recovers 
the best up-to-date SM predictions in absence of NP.

%%%%%%%%%%%%%%%%%%%%%%%%%%%%%%%%%%%%%%%%
\section{QED corrections for the $h\to 4\ell$ dilepton spectrum}
%%%%%%%%%%%%%%%%%%%%%%%%%%%%%%%%%%%%%%%%

In this section we describe how leading QED radiative corrections affect the dilepton 
spectrum of $h\to 4 \ell$ decays assuming a generic PO decomposition of the amplitude.
As anticipated,   we focus our discussion to 
the case of two lepton pairs with different flavor ($h\to 2e 2\mu$) and, more precisely, 
on the double differential  lepton-pair invariant-mass distribution 
\be
F(m_{ee},m_{\mu\mu})=\frac{\dd^2\Gamma (h\to 2e 2\mu) }{\dd m_{ee} \dd m_{\mu\mu} }.
\ee

The emission of soft and collinear photons leads to infrared (IR) divergences
in the $h\to 4 \ell$ spectrum. The full structure of such divergences is rather complicated. 
However, as we have checked by means of an explicit calculation at $O(\alpha)$,  
such divergences can be factorized in $F(m_{ee},m_{\mu\mu})$  
and can be analyzed separately for each dilepton system. This happens because 
each fermion current in Eq.~(\ref{eq:ThreePointFunction})  
carries an overall neutral electric charge. 

Working in the limit of massless leptons, we need to introduce two independent IR regulators
for soft and collinear divergences. We choose them to be:
i) the minimal fraction of invariant mass lost by the dilepton invariant-mass system;
ii) the minimal invariant mass of a single lepton plus (collinear) photon ($m^*$).
 
We then define the radiator $\omega(x, x_*)$, that represents the probability density function (PDF) 
that a dilepton system retains a fraction $\sqrt{x}$ of its original invariant mass after brems\-strah\-lung for a given $x_* \equiv  2 m_*^2/ m_0^2 $, where $m_0^2$ is the initial dilepton invariant mass (pre brems\-strahlung). 
By construction, the kinematical range of $x$ is
\be
0 < x  <  x_{\rm max} = 1 - x_* ~.
\ee
Keeping only the leading terms for $(1-x) \ll 1$ and $x_* \ll 1$, the radiator is
\be
	\omega(x,x_*) = \omega_1(x,x_*) \theta(1 - x_* - x) + \omega_2(x,x_*) \delta(1-x)~,
\ee
where
\be\begin{split}
	\omega_1(x,x_*) & = - \frac{\alpha}{\pi} \left( 1 + x - \frac{2}{1-x}\right) \log \left( \frac{2(1-x) - x_*}{x_*} \right)~, \\
	\omega_2(x,x_*) & = 1 + \frac{\alpha}{2\pi} \left[ \frac{\pi^2}{3} - \frac{7}{2} - 3 \log\left(\frac{x_*}{2}\right) - 2 \log\left(\frac{x_*}{2}\right)^2 \right]~.
\end{split}\ee
The first term, $\omega_1$, describes the real emission of a photon such that the lepton pair retains a fraction $\sqrt{x}$ of its invariant mass; the $\theta$-function implements the corresponding IR cutoff. 
The second term, $\omega_2$, 
describes the events in which the soft radiation is below the IR cutoff, as well as the effect of virtual corrections.

We have determined the  structure of $\omega_1$ by means of an explicit  $O(\alpha)$ calculation of the real emission,  
while $\omega_2$ has been determined by the condition $\int_0^1 \dd x \omega(x,x_*) =1$.
The latter condition implies a redefinition of $O(\alpha/\pi)$, not enhanced by large logs, 
of the PO characterizing the non-radiative amplitude.

Denoting by $m_{01}$ and $m_{02}$ the invariant masses of the two dilepton systems before bremsstrahlung, 
defining further $x_{i}=(m_{i} / m_{0i})^2$, it is easy to show that 
\begin{align} \label{eq: full diff}
\frac{\dd^4 \Gamma}{\dd m_{01} \dd m_{02} \dd x_1 \dd x_2}  = F_0(m_{01},m_{02}) \omega(x_1, x_{1*}) \omega(x_2, x_{2*})~,
\end{align}
where $F_0(m_{01},m_{02})$ denotes the non-radiative (tree-level) spectrum~\cite{Gonzalez-Alonso:2014eva}.

Starting from Eq.~(\ref{eq: full diff}) we can extract the 
double differential spectrum after radiative corrections. To this purpose,
we first trade $x_{1,2}$ for $m_{1,2}$, obtaining:
\ba 
\frac{\dd^4 \Gamma}{\dd m_{01} \dd m_{02} \dd m_1 \dd m_2}  &=& \frac{4 m_1 m_2}{m_{01}^2 m_{02}^2} F_0(m_{01},m_{02}) 
\nonumber \\
&& \!\! \times \omega\left(\frac{m_1^2}{m_{01}^2}, \frac{2m_*^2}{m_{01}^2}\right) \omega\left(\frac{m_2^2}{m_{02}^2}, \frac{2m_*^2}{m_{02}^2}\right).
\label{eq: full diff2}
\ea
From Eq.~(\ref{eq: full diff2}) we then explicitly extract the double differential decay width by integrating over all the possible physical $m_{01,02}$ combinations, determined by the conditions $m_{01}+m_{02}\leq m_h$ and $m_{01,02} \geq m_{1,2}/\sqrt{x_{\rm max}}$.
In this way we finally obtain: 
\ba
\!\!F(m_1,m_2) \!\!&=& \!\!\int_{\sqrt{m_1^2+2m_*^2}}^{m_h} \dd m_{01} \int_{\sqrt{m_2^2+2m_*^2}}^{m_h - m_{01}} \dd m_{02}   
~\frac{4 m_1 m_2}{m_{01}^2 m_{02}^2}  \nonumber \\
&& \!\!\! \times F_0(m_{01},m_{02}) 
 \omega\left(\frac{m_1^2}{m_{01}^2}, x_{1*}\right) \omega\left(\frac{m_2^2}{m_{02}^2}, x_{2*}\right).
 \label{eq:Ffinal}
\ea

We stress that the result in Eq.~(\ref{eq:Ffinal}) 
includes both real and virtual QED corrections. The latter have been indirectly determined by the normalization condition for $\omega(x,x_*)$,
that is the same condition applied in showering 
algorithms~\cite{Davidson:2010ew}. As anticipated, this implies a $O(\alpha/\pi)$ redefinition of the PO
compared to their tree-level values (both within and beyond the SM).
In the context of  next-to-leading order  (NLO) effective field theory 
(EFT) calculations~\cite{Hartmann:2015oia,Ghezzi:2015vva},
this procedure provides a well-defined condition for the matching between the full EFT calculation
of the amplitude and the PO decomposition.

%%%%%%%%%%%%%%%%%%%%%%%%%%%%%%%%%%%%%%%%
\section{Comparison with full NLO electroweak corrections}
%%%%%%%%%%%%%%%%%%%%%%%%%%%%%%%%%%%%%%%%

In this section we present a comparison of  the SM predictions for the $h\to 2e 2\mu$
dilepton invariant mass spectrum obtained using full NLO electroweak corrections~\cite{Bredenstein:2006rh},
and the PO decomposition ``dressed" with leading QED corrections, as described above. 

\begin{figure*}[t]
  \centering
  \begin{tabular}{cc}
   \hspace{-0.5cm} \includegraphics[width=0.5\textwidth]{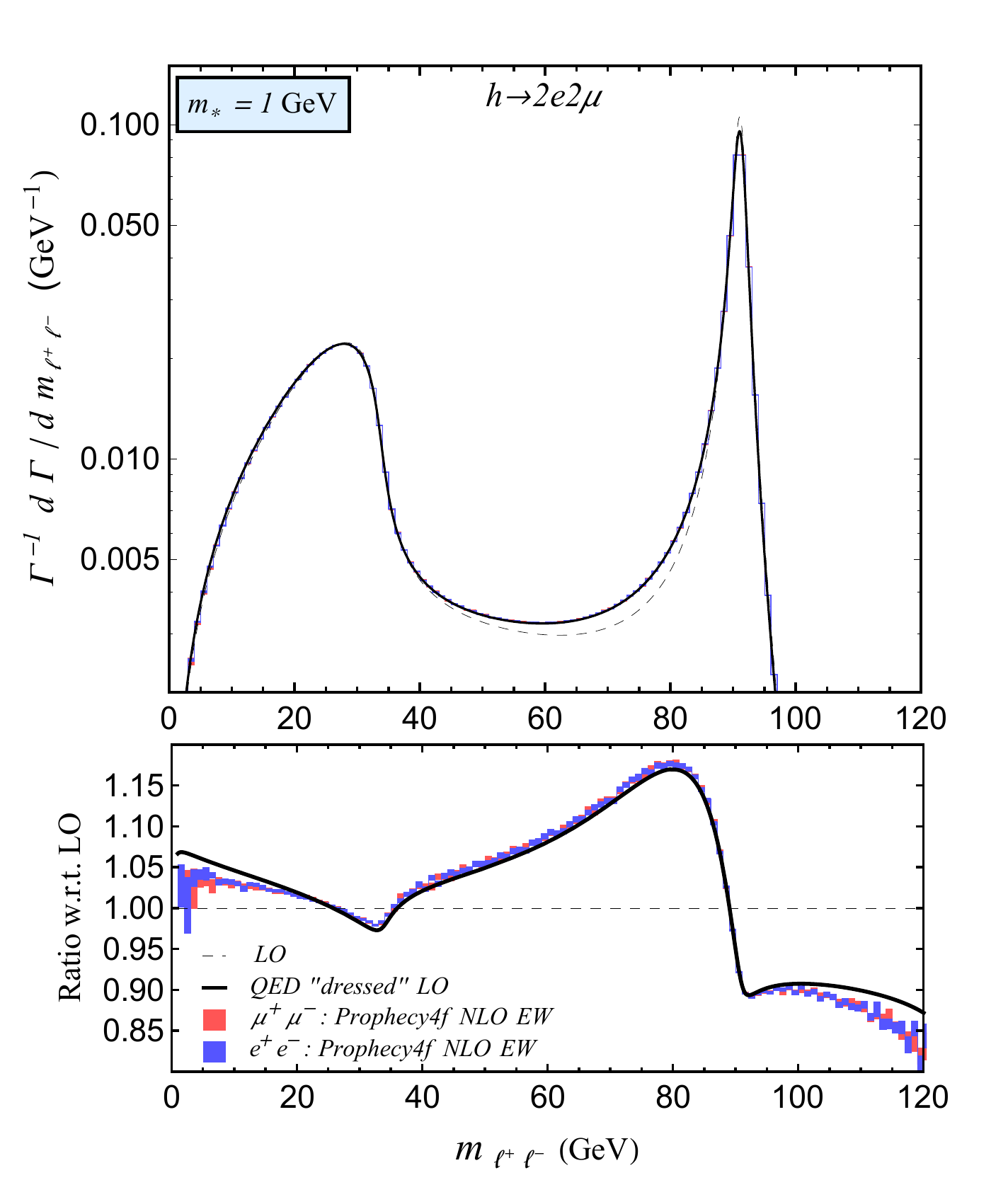} &
    \includegraphics[width=0.5\textwidth]{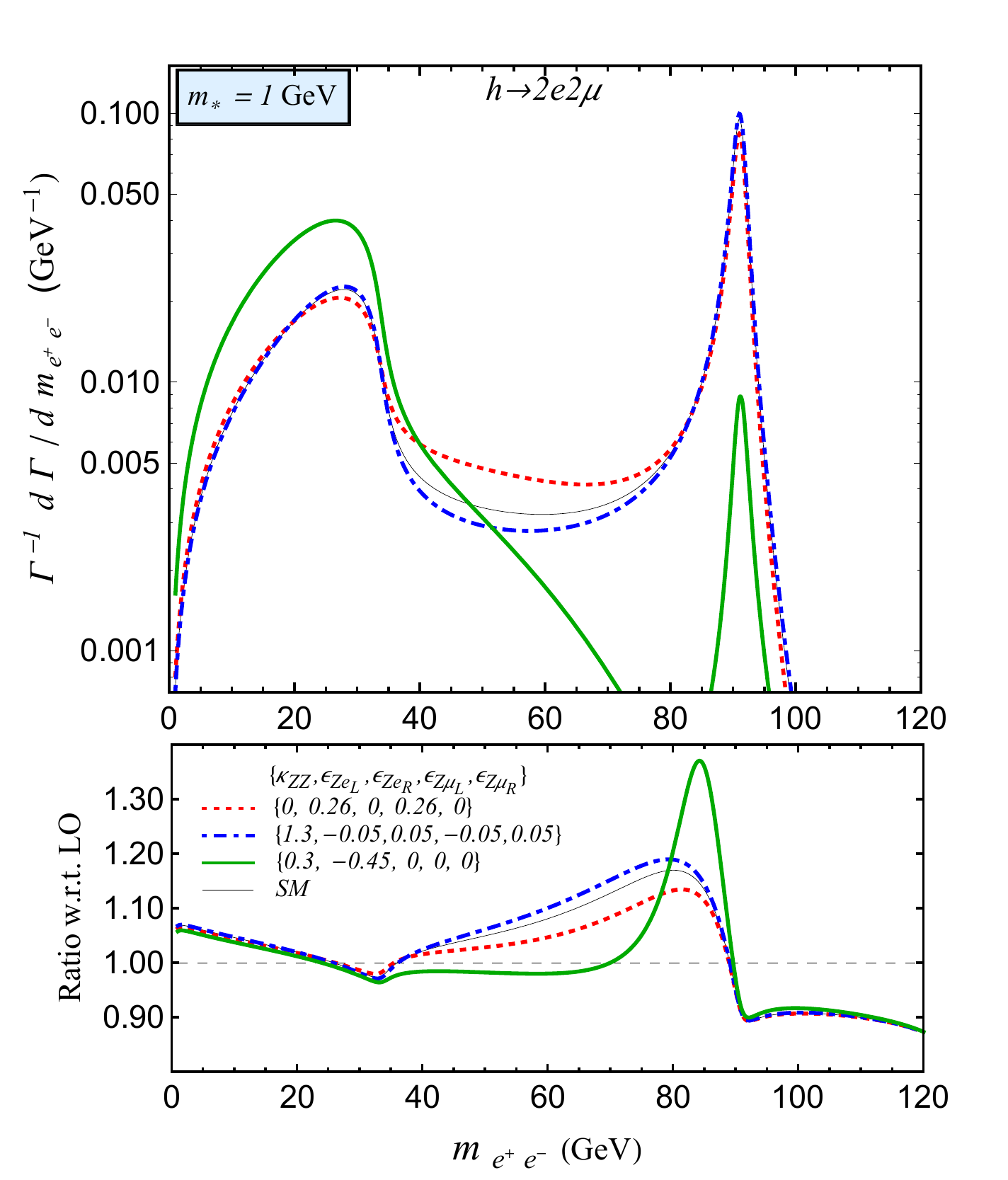} \\
     \end{tabular}
    \caption{ \small Left: Dilepton invariant mass spectrum in the SM for the $h\to 2e2\mu$ decay (full line: PO decomposition 
    ``dressed'' with QED corrections; red and blue bands: complete NLO result from Prophecy4f). Right: Dilepton invariant mass spectrum in the presence of new physics for various benchmark scenarios (see text for details). \label{fig-prva}}
\end{figure*}

The complete SM NLO electroweak corrections to $h\to4\ell$ 
have been computed in~\cite{Bredenstein:2006rh}, and the results have been implemented in the Monte Carlo event generator 
Prophecy4f~\cite{prophecy4f}. 
We have used Prophecy4f version 2.0 to generate $200$ millions weighted events for the recombination 
mass parameter  $m_* =1$~GeV. We have used the default Prophecy4f SM inputs 
except for setting the Higgs boson mass to $125$~GeV.  Prophecy4f adopts the dipole subtraction formalism~\cite{Dittmaier:1999mb}
for the treatment of soft and collinear divergences, and the so-called ``photon-recombination" is applied. 
In particular, if the invariant mass of a lepton and a photon is smaller than $m_*$, the 
photon momentum is added to the lepton momentum~\cite{Bredenstein:2006rh} .
As a result, $m_*$ coincides with the collinear cut-off introduced in the previous section.

In Fig.~\ref{fig-prva} (left) we show the decay distribution as a function of the dilepton invariant mass normalized to the 
total decay width for $h\to 2e2\mu$ in the SM (upper plot) and the ratio between NLO and leading-order (LO) predictions (lower plot). 
Shown in solid black is our improved prediction obtained by convoluting the leading order distribution, shown in dashed black, with the radiator function as described in the previous section. The PO have been fixed to their SM tree-level reference values ($\kappa_{ZZ}=1$, $\epsilon_i=0$~\cite{Gonzalez-Alonso:2014eva}).  
The Prophecy4f predictions within MC uncertainty are shown with red and blue bands for $\mu^+\mu^-$ and $e^+ e^-$ invariant mass 
spectra, respectively. 

We list here a series of conclusions that can be derived from this numerical comparison.
\begin{itemize}
\item{} 
The spectrum obtained with the PO decomposition of the amplitude, ``dressed" with leading QED corrections, provides an excellent 
approximation (within 1\% accuracy) to the spectrum obtained with full NLO EW corrections.\footnote{The $\sim 2\%$ deviations 
at the border of the phase space are expected due the breakdown of the approximation $m_{\ell\ell} \gg m_*$ 
employed in the analytic evaluation of the radiation function.}
\item{}  
The effect of the leading QED corrections can be large, exceeding $10\%$ in specific regions of the phase space.
It therefore must be included, in view of a precise data-theory comparison, also when fitting beyond-the-SM parameters.
\item{}  
The  PO   ``dressed" spectrum is obtained setting $\epsilon_i=0$  
(i.e.~to their LO SM values). The good agreement with the complete NLO calculation confirms that the $O(\alpha/\pi)$ redefinition of the 
$\epsilon_i$ is a small effect, with no observable consequences for the $h\to 2e2\mu$  dilepton
invariant mass spectrum.
\end{itemize}

%%%%%%%%%%%%%%%%%%%%%%%%%%%%%%%%%%%%%%%%
\section{Implications for New Physics}
%%%%%%%%%%%%%%%%%%%%%%%%%%%%%%%%%%%%%%%%

As shown in Fig.~\ref{fig-prva} (left), 
radiative corrections can be sizable and must be included also when going 
beyond the SM. Having demonstrated the validity of our QED improved predictions
to describe such effects, we are in position to apply the method in the presence of an arbitrary new physics 
contribution to $h\to 2e 2\mu$ decay as parameterised by generic PO~\cite{Gonzalez-Alonso:2014eva}.
As an illustrative example, we consider the impact of the leading QED corrections
for non-standard values of $\kappa_{ZZ}$, $\epsilon_{Ze_L}$, $\epsilon_{Ze_R}$, $\epsilon_{Z\mu_L}$, and $\epsilon_{Z\mu_R}$. 

To draw some general conclusions we analyse  three ben\-ch\-mark points, chosen such that the deviations of the total $h\to 2e 2\mu$ decay rate from the SM prediction are always small,\footnote{The dependence of the total rate on the PO can be found in Ref.~\cite{Gonzalez-Alonso:2015bha}.} but the impact on the spectrum are quite different. 
 The results of the inclusion of QED corrections are shown in Fig.~\ref{fig-prva} (right). As 
in the left panels,  we plot the dilepton invariant mass distribution normalized to the total rate (upper plot) 
and the ratio  between NLO and LO (lower plot). 

The definition of the benchmarks, and the consequences following from the analysis of radiative 
corrections,  are listed below.
\begin{itemize}
\item {\bf Benchmark I} [$\kappa_{ZZ}=1.3$, $\epsilon_{Ze_L}=\epsilon_{Z\mu_L}=-0.05$ , $\epsilon_{Ze_R}=\epsilon_{Z\mu_R}=0.05$   ({\em dot-dashed blue})].\\
Here the deviation from the SM point in the Higgs PO parameter space is small: this benchmark point is  compatible with naive 
power counting in the linear EFT expansion.  As a consequence, small deformations in the spectrum are obtained (upper panel) and the relative QED corrections are SM-like (lower panel). In this regime, the leading QED corrections can be directly extracted from
the SM result (via an appropriate NLO/LO re-weighting).

\item  {\bf Benchmark II} [$\kappa_{ZZ}=0$, $\epsilon_{Ze_L}=\epsilon_{Z\mu_L}=0.26$, $\epsilon_{Ze_R}=\epsilon_{Z\mu_R}=0$
({\em dotted red})]. \\
Here the deviation from the SM point is sizable, beyond the naive power counting within a generic 
EFT (both linear and non-linear). However, the PO configuration is such that the deviations from the SM in the spectrum are small.
This implies that the relative impact of QED corrections is  still SM-like.

\item {\bf Benchmark III} [$\kappa_{ZZ}=0.3$, $\epsilon_{Ze_L}=-0.45$ and $\epsilon_{Z\mu_L}=\epsilon_{Ze_R}=\epsilon_{Z\mu_R}=0$
({\em solid green})]. \\
In this example we observe a sizable distortion of the dilepton shape (upper panel). As a consequence, the relative 
impact of the QED corrections is quite different from the SM case (a description of radiative corrections by 
NLO/LO re-weighting of the SM result would not provide a good approximation).

\end{itemize}

%%%%%%%%%%%%%%%%%%%%%%%%%%%%%%%%%%%%%%%%
\section{Conclusions}
%%%%%%%%%%%%%%%%%%%%%%%%%%%%%%%%%%%%%%%%

The dominant electroweak corrections to $h\to 4\ell$ decays are due to the universal 
soft and collinear photon emission. As shown in Fig.~\ref{fig-prva}, these 
can lead to distortions of the dilepton invariant  spectrum of  
 $O(10\%)$ is specific regions of the phase space. These effects  are of the same order
as the expected modifications from the SM under the assumption of underlining linear 
EFT~\cite{Gonzalez-Alonso:2015bha}. It is then mandatory 
to properly incorporate these corrections in a consistent way both within and beyond the SM.

As we have shown in this paper, this can be achieved in general terms within the 
framework of the Higgs PO.  In particular we have shown that: (i) the QED corrected
predictions for the $h\to2e2\mu$ dilepton invariant mass spectra, with PO fixed to their SM LO values,
are in agreement with the full NLO electroweak SM predictions within 
$1\%$ accuracy; (ii)~the QED corrections in the presence of NP can be sizable and significantly 
different from the SM case.

\begin{acknowledgements}
This research was supported in part by the Swiss National Science Foundation 
(SNF) under contract 200021-159720.
\end{acknowledgements}

%%%%%%%%%%%%%%%%%
\end{document}